# Can the fluctuations of the motion be used to estimate performance of kayak paddlers?

**Gergely Vadai[1] and Zoltán Gingl[1]**

[1]Department of Technical Informatics, University of Szeged, Árpád tér 2, 6720 Szeged, Hungary

e-mail: vadaig@inf.u-szeged.hu

**Abstract.** Today many compact and efficient on-water data acquisition units help the modern coaching by measuring and analyzing various inertial signals during kayaking. One of the most challenging problems is how these signals can be used to estimate performance and to develop the technique. Recently we have introduced indicators based on the fluctuations of the inertial signals as promising additions to the existing parameters. In this work we report about our more detailed analysis, compare new indicators and discuss the possible advantages of the applied methods. Our primary aim is to draw the attention to several exciting and inspiring open problems and to initiate further research even in several related multidisciplinary fields. More detailed information can be found on a dedicated web page, http://www.noise.inf.u-szeged.hu/kayak.

## 1. Introduction

Periodic processes are very common in various disciplines. The processes can be inherently periodic with an intrinsic rate, like the heart function, or can be driven periodically as the ambient temperature on the surface of the Earth. Artificial systems including machines often have periodically moving parts, motors also. In several cases deterministic or random changes of the operation frequency, noise in the period can be informative about the proper operation of the system, can be a good measure of quality, indicator of dysfunction or even predictor of a possible forthcoming damage. Noise coming from such a system can be an efficient diagnostic tool in both inherently periodic or in periodically driven systems. The wide range of examples include the analysis of heart rate variability [1], hemodynamic regulation during metronomic breathing [2], gait dynamics, fluctuations of human walking [3], daily activity measured by actigraphs, smart watches [4-6], daily temperature and other environmental fluctuations [7,8], period fluctuations of variable stars [9], fault diagnosis of induction motors [10-11]. Note that having noisier period is not necessarily bad – too small heart rate variability can indicate a possible heart disease [1]. Oddly enough noise can even play constructive role: adding noise can improve signal to noise ratio via the mechanism of stochastic resonance [12-14].
Following the idea of using fluctuations as a diagnostic tool we have introduced noise analysis methods to estimate the performance of kayak paddling [15]. Inertial sensors like accelerometers and gyroscopes are used in coaching devices developed for professional kayak paddlers and trainers [16-19], and the measured quantities and their changes in a stroke cycle are used to classify the athletes' performance [20-23]. In our previous work we have suggested that – since the optimal motion of a kayak can be assumed to be purely periodic – the fluctuations of its period could be an indicator of the quality of paddling [15]. We have calculated time and frequency domain parameters and we have introduced a promising signal-to-noise ratio (SNR) estimation using the raw signals without the typically required detection of strokes.

In this paper we report on our latest results presented at the conference of Unsolved Problems of Noise (UPON) in an invited talk [24]. We show a more detailed analysis of the introduced older and new indicators both in the time and frequency domain. Following the spirit of the conference here we focus on the most interesting open questions that can be inspiring not only for the noise research community but for a wider range of scientific and engineering audience also. It is important to note that the results can be related to many multidisciplinary applications as well.

## 2. Kayak motion data

The motion signals of the kayaks were measured by a special portable instrument developed in our laboratory for this purpose [25]. The device contains a 3-axis accelerometer and a 3-axis gyroscope to support acquisition of the most important inertial signals, see figure 1. The built in microcontroller's data converter digitizes these signal with a sample rate of 1000 Hz.

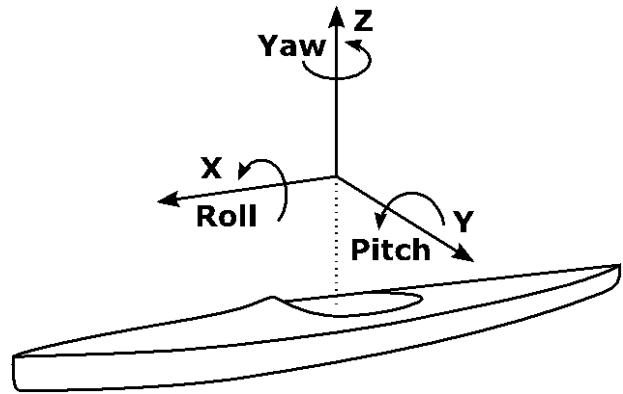

**Figure 1.** Orientation of the 3-axis accelerometer and 3-axis gyroscope mounted in the kayak.

Since there can be several different types of paddlings (for example due to different tasks at trainings or at races), one can have different aims of the data analysis like optimizing paddling techniques, detecting faults, examining long-term evolution of technical parameters, comparing paddlings at races, etc. Our primary aim was the general estimation of the athletes' performance using fluctuation analysis based methods that could be very useful at many of the mentioned purposes. In order to examine indicators for performance estimation the really important question is: can we classify the performance or technical skills? In the case of different athletes' paddling, their age could be used like in our related works [15, 25], but its connection with skills is not always clear. In this paper we used classification done by the trainer in the scale: 1-10, too.

Another problem is to find how we can compare paddlings of several athletes using different paddling techniques especially if they are influenced by significantly different conditions. In order to compare the typical performance of paddlers in very similar circumstances we have analysed the first 10 minutes of long range (>5 km) training paddlings of 14 athletes with different age and technical skills.

Note that another approach can be the analysis of one athlete's different paddlings for example testing the indicators in the function of race times. This analysis needs systematic measurements at many trainings and real or simulated races.

As it is shown on figure 2, the fluctuation-based indicators were calculated (both in the time and frequency domain) for shorter time window widths (30 seconds), and the averages for the examined 10 minutes long part were compared. Note that in the case of spectral methods all six inertial signals were used instead of using the x-axis acceleration only.

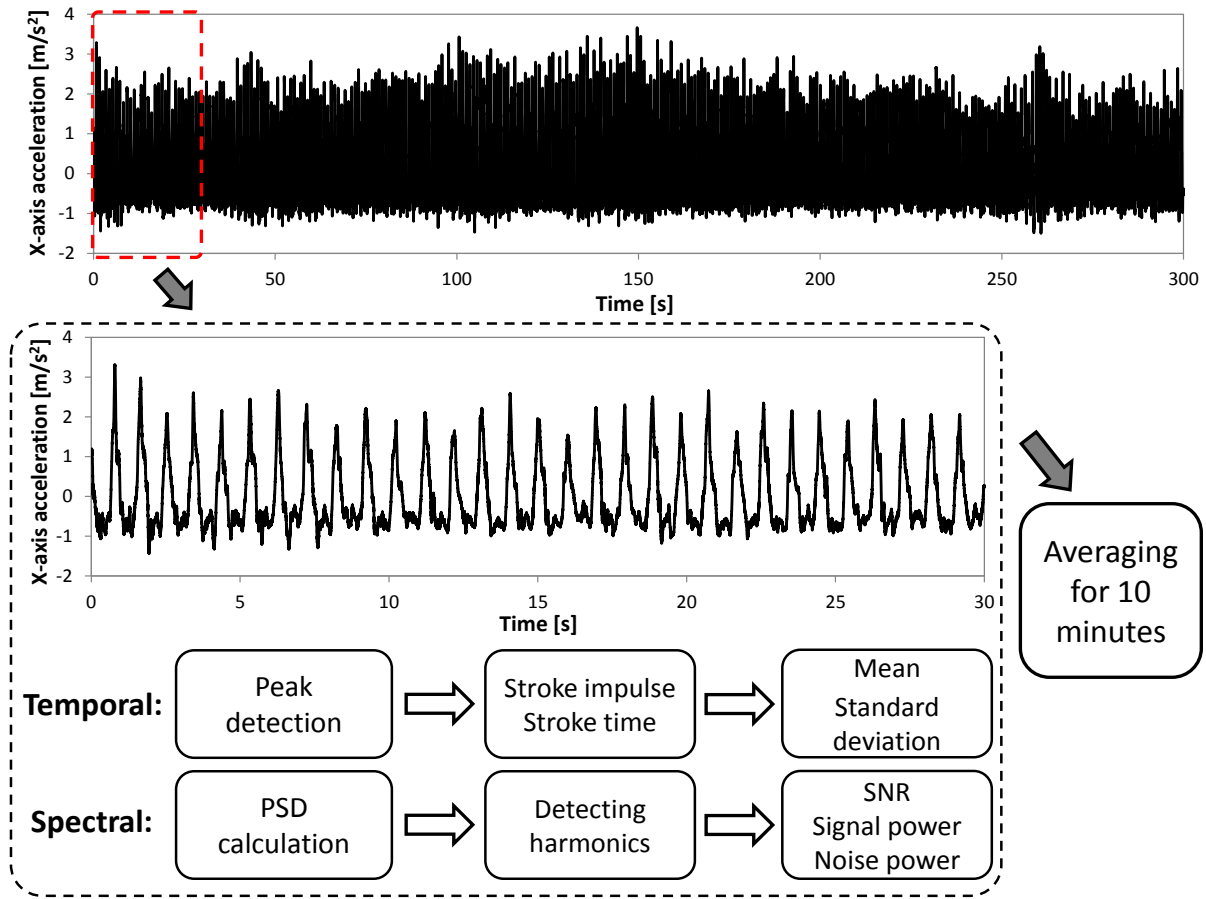

**Figure 2.** Calculation of the fluctuation-based indicators both in the time and frequency domain.

## 3. Temporal indicators

In the time-domain the forward axis (x-axis) acceleration plays the major role in the analysis of the motion. The interpretation of the signals' shape and its connection with an optimal stroke cycle were discussed in several papers [19-23]. Furthermore the classical parameters of a stroke cycle could be calculated on this signal after identifying each stroke using peak and level crossing detection algorithms. The most important quantities, which we use below, are illustrated on figure 3. The stroke cycle is characterized by its time length (stroke time) and the speed increase in the pulling phase (stroke impulse) which can be calculated by integration of the positive part of the acceleration signal [25]. The parameters were also calculated for the total duration of the left and right hand stroke, which is the total period of the motion (two hands stroke time).

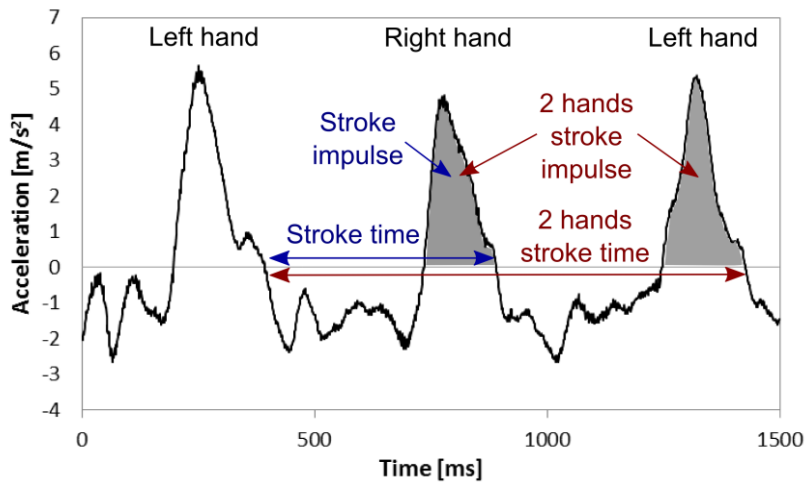

**Figure 3.** Classical parameters of a stroke cycle detected on x-axis acceleration signal.

The mean values and trend curves of these classical parameters or the shape of the raw signals can be helpful for trainers and athletes. Unfortunately the analysis is rather complex and time-consuming in most cases. Easily usable but still accurate and reliable indicators of performance or technical skills could be useful at many levels of the trainers' work.

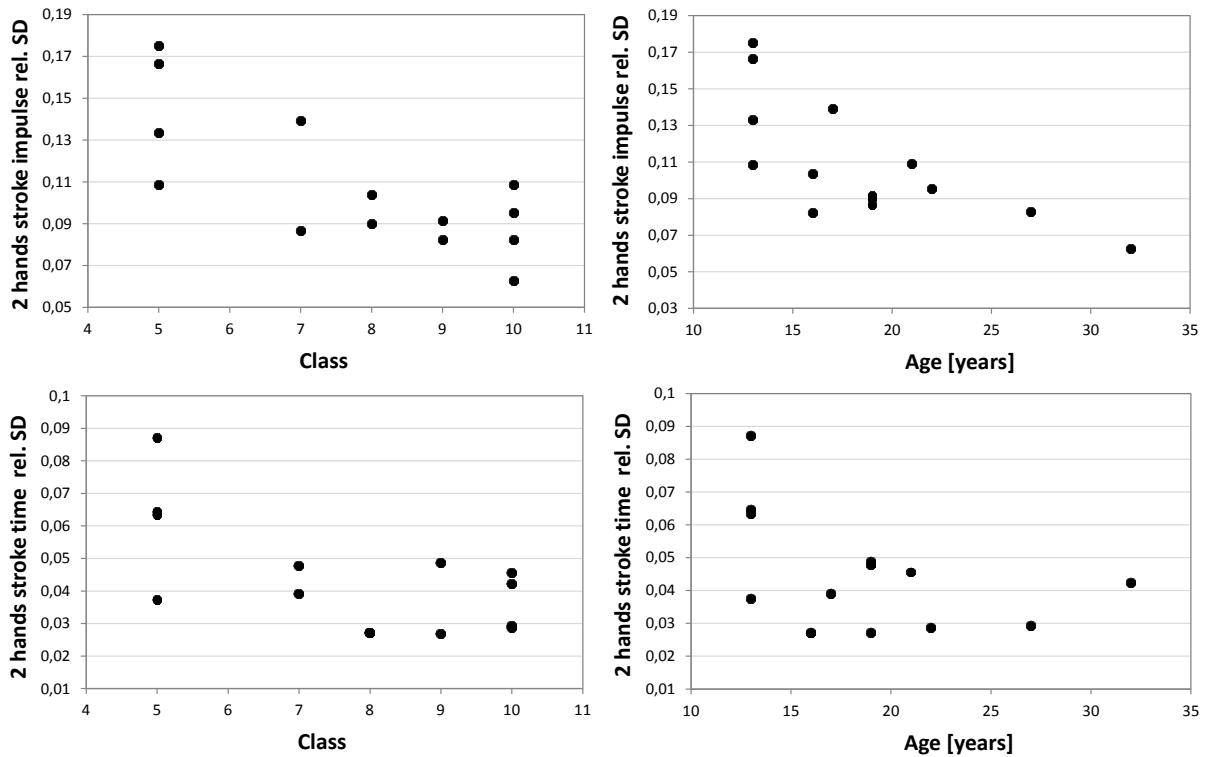

**Figure 4.** Relative standard deviation of the two hands stroke impulse and two hands stroke time in the function of the technical skill's classification (left) and in the function of athlete's age (right).

The idea behind using fluctuation analysis is to measure the variability of kayaks periodic motion because the steadiness of the rate could be correlated with the quality of the paddling [15]. This correlation is shown on figure 4, where the relative standard deviation (SD) of the two hands stroke impulse and the two hands stroke time decreases with better class and age significantly. Each point on these plots corresponds to an athlete and was defined as it is shown on figure 2: the standard deviation parameters (SD) were calculated for each 30 seconds wide time windows, and were averaged over the first 10 minutes of a long-range paddling at training.

Neither the athletes' age nor the classification indicates exactly their technical skills, nevertheless the relationship between performance and variability seems to be evident. On the other hand, there are some open questions about how one should calculate these SD-s. On figure 5, we compared SD-s of the stroke impulse calculated in different ways in the function of classes using the coefficient of determination ($R^2$). As it can be seen, the relative SD-s shows better correlation than the absolute SD-s.

Changing stroke rate and effects of tiredness can be observed in every paddling, so the length of the processed data and the use of detrending algorithms can have some impact on the indicators' values and their observed relationship with technical skills. As one can see on figure 5, in the case of the 30 seconds long evaluation the detrending has no significant role, however in the case of comparing long-length race paddlings it can be useful.

The presented methods can be used to analyse the periodicity of the kayak's motion. In the case of the forward axis acceleration, one can consider the period as one stroke cycle or the duration of a left and a right hand stroke, too. Therefore it is a really important question if indicators related to one hand or two hands have stronger relationship with performance. As depicted on figure 5, the two hands stroke impulses SD shows better correlation with classification in every case.

Comparison of the SD-s of the stroke time shows the same ratio between $R^2$ of different methods but the coefficients has lower values which fact is consistent with trends shown on figure 4. The reason for the relative low $R^2$ values is the low number of data points used for calculating the regression, but we note that all SD-s show better correlation than the mean value of the stroke impulse that has already known relationship with performance.

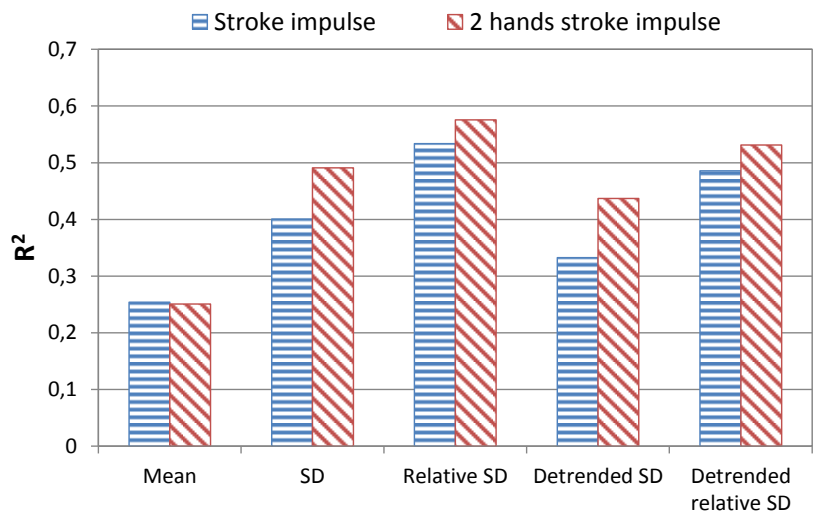

**Figure 5.** Comparison of the SD-s of stroke impulse and two hands stroke impulse calculated in different ways in the function of classes using the coefficient of determination. Relative SD was defined as the ratio of the absolute SD and the mean. The data and steps of signal processing are described in section 2.

## 4. Spectral indicators

Detecting the strokes using the complex signals with additional irregularities and noise can be rather complicated and inaccurate in most cases. Uncertainty of time-domain peak and zero crossing detection can be eliminated by using indicators calculated in the frequency domain. In this case the power spectra of the raw signals can be used to derive indicators to describe the period fluctuations. Following this idea we have introduced a certain kind of signal-to-noise ratio (SNR) as another possible indicator of performance [15]. Note that it can be even extended to a joint time-frequency analysis that allows monitoring of the time dependence of the spectral indicators too.

One of the most important questions is how one can separate the "signal" and the "noise" in the power density spectra. In the case of the forward axis acceleration and the yaw and roll gyroscope signals, the dominant frequency is the first harmonic that belongs to the one hand stroke cycle. On the other hand, in the case of the other three signals, the fundamental frequency is more significant as these signals belong to the whole period of both hands strokes. Figure 6 shows examples of these two cases. The first harmonic is the dominant peak in the x-axis acceleration power spectral density (PSD), but the dominant frequency of roll axis angular velocity is the fundamental frequency. Furthermore, the magnitude of the harmonics in the two cases are different, therefore calculating the indicators can differ significantly.

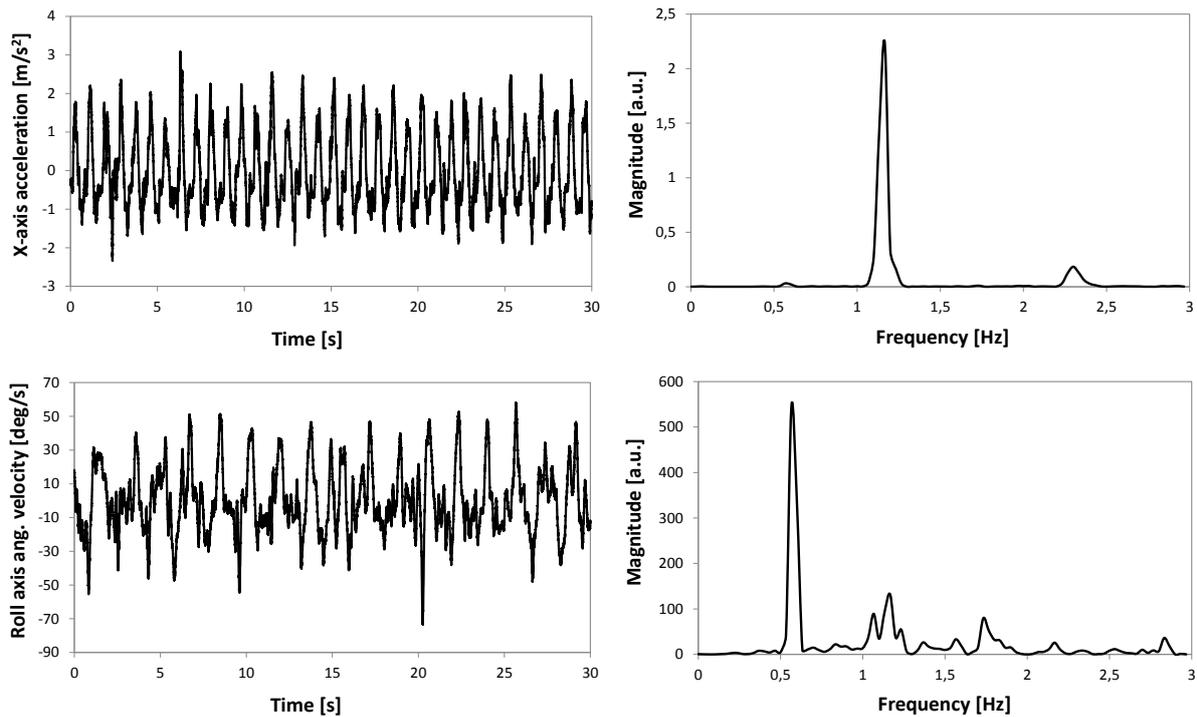

**Figure 6.** X-axis acceleration and roll axis angular velocity signal and its PSDs. The two signals have different dominant frequencies: first harmonic for the x-axis acceleration and the fundamental frequency for the roll axis angular velocity.

In order to calculate the SNR we consider the area under the harmonic peaks as the signal power and the rest of the power as noise. Note also that there are several ways to calculate these values and therefore the SNR. It can depend on the number of harmonics taken into account for the signal part and the area under the peaks is a function of the extent used in the calculations. Besides using the SNR, both of its components can be used as indicators, too. Figure 7 shows a certain type of SNR (details can be found in its caption) calculated for the roll axis gyroscope signal. The strong relationship with both the technical quality and with the paddlers' age can be clearly seen.

Since several paddling techniques can lead to high performance, the corresponding signal power is not necessarily a good indicator – the magnitude of the harmonics can be rather different. However the noise power seems to reflect the technical skills much better. These facts are illustrated on Figure 8.

The spectral indicators were obtained using the same signals and time windows as in the case of the temporal indicators. Namely the spectral indicators were calculated over 30 second long windows, and were averaged for the first 10 minutes of a long-range paddling at training.

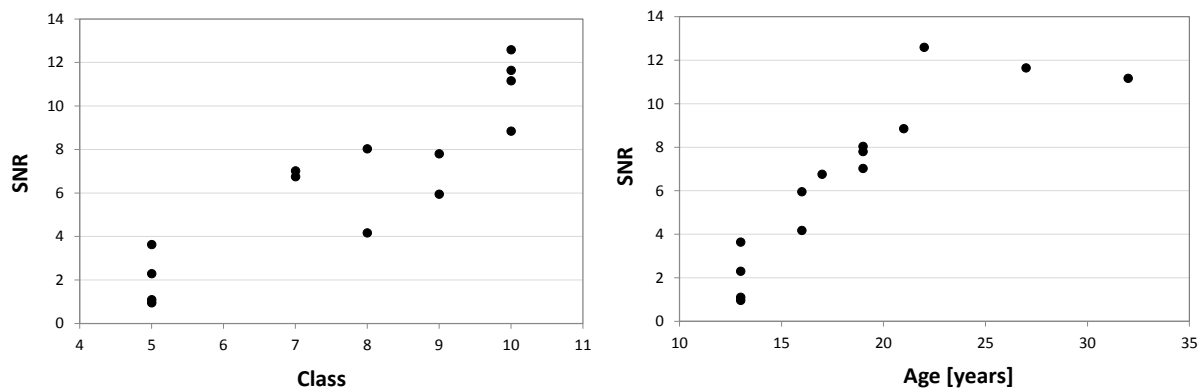

**Figure 7.** Signal to noise ratio (SNR) of roll axis angular velocity as a function of technical skills' classification (left) and of athletes age (right). Signal power of the spectrum were defined as the power of 0.2 Hz width peaks on fundamental frequency and first 5 harmonic frequencies, while the noise power was the power of the rest of the spectrum.

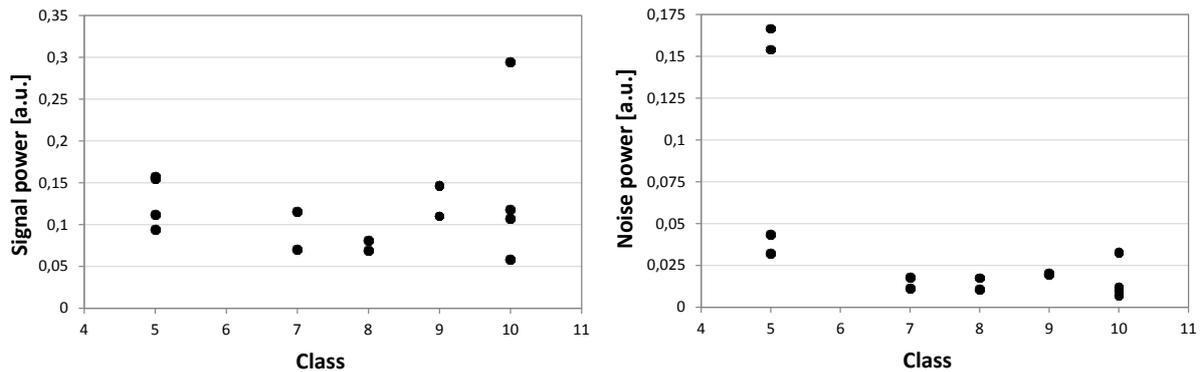

**Figure 8.** Signal and noise power of roll axis angular velocity as a function of technical skills' classification. Same method was used as at figure 7.

Figure 9 depicts the coefficient of determination between the class and SNR, signal and noise power for all six inertial signals. As we have pointed out above, the SNR appears to have the strongest relationship with technical skills for almost all inertial signals. It can be seen that the best correlation with the class is obtained for the roll and yaw axis angular velocities. For these two signals the dominant frequency is equal to the fundamental frequency, therefore one can conclude that the two hands spectral indicators characterize the performance better.

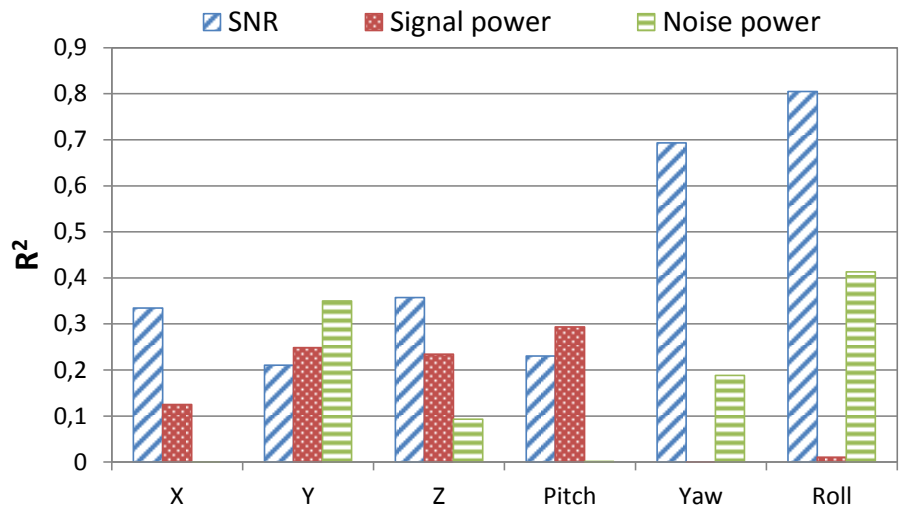

**Figure 9.** Comparison of three spectral indicators calculated for six motion signals in the function of classes using the coefficient of determination. The data and steps of calculating PSDs is described in section 2. At the calculation of indicators, same method was used as at figure 7.

We have compared two signal and noise definitions by using two different numbers of harmonics for the calculations. As it is depicted on both plots in figure 10, SNR values where the first 6 peaks of the spectrum were defined as signal (fundamental frequency and the first five harmonics) describe the technical skills much better than SNR using only the first two peaks (frequency of one hands and two hands stroke).

Detecting the peak location and extent in the spectrum accurately can be a problem, however on the other hand it is desirable to find simple and universal methods that can be used for all kinds of signals and paddling techniques. We have designed and tested numerical methods for finding the signal power in the spectra based on fixed peak width of 0.2 Hz and estimated half-width. We have used these in the above mentioned different signal and noise definitions based on six and two harmonics. We have compared all of these results for two different spectral window types (rectangular and Hanning). We have found no significant difference between integrating the peaks over a predefined peak width or over a frequency dependent half-width based extent. The results are also insensitive to the choice of the window functions we have tested. The obtained coefficients of determination are plotted on figure 10.

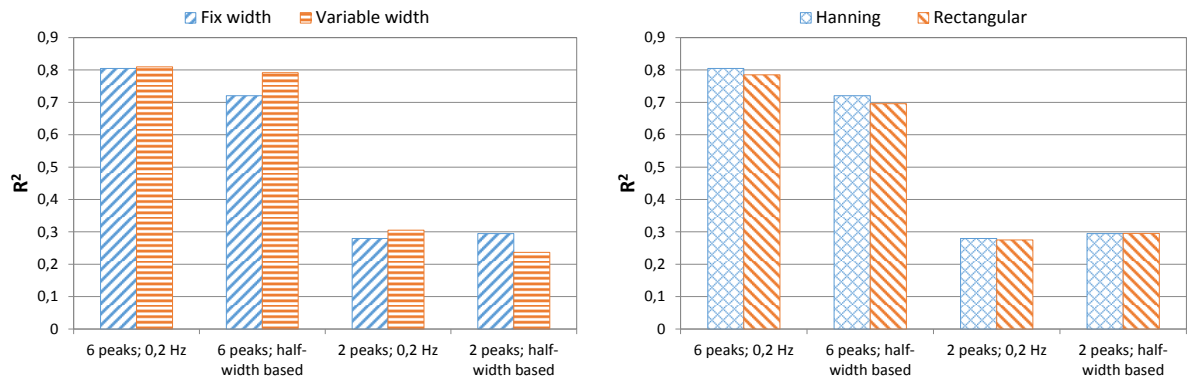

**Figure 10.** Comparison of fix and widening peak-width based methods (left graph) and comparison of Hanning and rectangular windows at PSD calculation (right graph). In the x-axis, different SNR definitions and PSD peak detect methods were compared, too. The data and steps of calculating PSDs is described in section 2.

Note that as in the case of temporal indicators, using different window lengths can have impact on the values of the indicators and their relationship with performance.

## 5. Conclusion and open problems

We have shown that noise analysis can be a promising diagnostic tool for the estimation of the performance of kayak paddlers. In addition to our previous results we have introduced and evaluated new time and frequency domain indicators. We have found that the most useful indicator of the quality is the SNR of the roll angular velocity and since it is calculated in the frequency domain, no complex and rather uncertain peak and level crossing detection is needed. We have investigated several different calculations of SNR, signal and noise power and developed simple algorithms to calculate these for six inertial sensor signals. We have found that both the spectral and time domain indicators worked well only for the signals whose dominant frequency is the fundamental frequency, when the period was the sum of a left and a right hand stroke.

Although it sounds likely that the steadiness of the motion has a primary role in the paddling quality, many open questions may arise. The following interesting questions and problems can be evidently identified.

- There can be other temporal parameters or spectral methods, indicators that can indicate the performance even better, so more detailed analysis could be useful.
- The indicators discussed above were tested using classification of the technical skills. However the actual performance of the athlete depends on many factors. This is exceptionally important in order to determine how reliably the indicators can be used for certain cases, to determine what kind of data processing is needed. There can be several problems, subjective elements about this.
- It is one of the most exciting questions what are the sources of the noise found in the paddling periodicity and strength? It can depend on mechanical effects – movement of the kayak and of the human body, dissipation –, learned technical skills and even mental condition.
- It is well known that mental condition can be a significant factor of excellent performance. However it is not clear, how it can affect the noise level, SNR or other indicators.
- The above mentioned open problem can be a starting point of neurology related experiments and analysis.
- We have tested a few methods of separating signal and noise; however it is not straightforward. It is not easy to define what should be considered as noise, what are the sources of noise and if they can be separated from each other. Slow – possibly randomly changing – drifts, lower and higher frequency components can appear easily.
- Although noise analysis is used in several diagnostic applications, it is not yet clear, how this fluctuation analysis can be related to other periodic motions and especially to other sport fields including running, swimming, cycling.

- Simulations can be very useful to know more about the processes, to find out the sources of noise, to test possible indicators and evaluation algorithms. It can also be important to support development of theoretical models and to verify their compliance.
- Since smart phones, smart watches have more and more integrated sensors including inertial sensors it seems to be possible to implement noise analysis algorithms also to evaluate health indicators, and use these during commercial sport exercises.

Future research may focus on answering these questions and the methods can have potential applications in many other fields as well.

## Acknowledgments

The authors thank Gergely Makan, Róbert Mingesz, János Mellár and athletes, trainers for their help and valuable discussions.

The publication/presentation is supported by the European Union and co-funded by the European Social Fund. Project title: "Telemedicine-focused research activities on the field of Mathematics, Informatics and Medical sciences" Project number: TÁMOP-4.2.2.A-11/1/KONV-2012-0073.